\documentclass[]{aastex631}

\usepackage{amsmath}
\usepackage{xcolor}

\begin{document}

\title{Detection of kink oscillations in solar coronal loops by a CNN-LSTM neural network}

\correspondingauthor{Sergey A. Belov}
\email{Sergey.Belov@warwick.ac.uk}

\author[0000-0002-3505-9542]{Sergey A. Belov}
\affiliation{Centre for Fusion, Space and Astrophysics, Department of Physics, University of Warwick, Coventry CV4 7AL, UK}

\author[0000-0002-1203-094X]{Yu Zhong}
\affiliation{Centre for Fusion, Space and Astrophysics, Department of Physics, University of Warwick, Coventry CV4 7AL, UK}

\author[0000-0002-0687-6172]{Dmitrii Y. Kolotkov}
\affiliation{Centre for Fusion, Space and Astrophysics, Department of Physics, University of Warwick, Coventry CV4 7AL, UK}
\affiliation{Engineering Research Institute \lq\lq Ventspils International Radio Astronomy Centre (VIRAC) \rq\rq, Ventspils University of Applied Sciences, Ventspils, LV-3601, Latvia}

\author[0000-0001-6423-8286]{Valery M. Nakariakov}
\affiliation{Centre for Fusion, Space and Astrophysics, Department of Physics, University of Warwick, Coventry CV4 7AL, UK}
\affiliation{Engineering Research Institute \lq\lq Ventspils International Radio Astronomy Centre (VIRAC) \rq\rq, Ventspils University of Applied Sciences, Ventspils, LV-3601, Latvia}
\affiliation{Centro de Investigacion en Astronom\'ia, Universidad Bernardo O'Higgins, Avenida Viel 1497, Santiago,Chile}

\begin{abstract}
A hybrid machine learning model which combines a shallow convolutional neural network and a long short-term memory network (CNN–LSTM), has been developed to automate the detection of kink oscillations in coronal plasma loops within large volumes of high-cadence sequences of imaging data. The network was trained on a set of 10,000 synthetic data cubes designed to mimic sequences of coronal images, achieving an accuracy greater than 98\% on this synthetic dataset. The model was then applied to detect kink oscillations in real data cubes of coronal active regions observed with SDO/AIA in the 171~\AA\ channel. This dataset consisted of 50 samples with visually detected kink oscillations and 128 samples without. Each sample covered an area of 260$\times$260~pixels in the spatial domain and a duration of 30~min with a 12~s cadence in the time domain. Both off-limb and {on-disk} regions of interest were used. The data were pre-processed by median filtering in the time domain, and Gaussian {smoothing and Contrast} Limited Adaptive Histogram Equalization in the spatial domain. In the real dataset, the performance of the model was 83.7\%. 
{The model is fully available in open access.} We regard the CNN–LSTM model developed as a first step toward creating robust tools for routine solar coronal data mining in the context of coronal oscillation study.
\end{abstract}

\keywords{Solar coronal loops --- Solar coronal waves --- Neural networks}

\section{Introduction} \label{sec:intro}

The interest in magnetohydrodynamic (MHD) wave processes in the corona of the Sun is motivated mainly by the coronal heating problem \citep[e.g.,][]{2020SSRv..216..140V} and seismological diagnostics of coronal plasma \citep[e.g.,][]{2024RvMPP...8...19N}. One of the most studied coronal wave phenomena is the kink oscillations of coronal plasma loops \citep[e.g.,][]{1999ApJ...520..880A, 1999Sci...285..862N}. The oscillations are transverse displacements of the bright loops in the plane of the sky, resembling oscillations of a string with fixed ends.  Kink oscillations of coronal loops are commonly detected with high-resolution EUV imagers \citep[see, e.g.,][for a comprehensive review]{2021SSRv..217...73N}. In addition, kink oscillations can manifest as periodic Doppler shifts in coronal emission lines \citep[e.g.,][]{2012ApJ...759..144T}, and modulate radio emission produced by flares \citep[e.g.,][]{2013SoPh..284..559K}. Kink oscillation periods typically range from about one minute to half an hour and are found to increase with the length of the oscillating loop \citep{2019ApJS..241...31N}. Typical displacement amplitudes reach several tens of Mm, but in all observed events are much smaller than the loop length. In the majority of the observed cases, a fundamental {transverse} standing harmonic is detected, with the maximum displacement at the loop top and nodes at its footpoints. In some cases, {higher axial harmonics} have been identified too \citep[e.g.,][]{2007ApJ...664.1210D}. 

Kink oscillations appear in two regimes, large-amplitude rapidly decaying oscillations and small-amplitude decayless oscillations. In both regimes, oscillation cycles appear to be harmonic. Decaying oscillations are usually excited by an initial displacement of the loop from an equilibrium by a low coronal eruption \citep{2015A&A...577A...4Z}, and rapidly {decay} within a few oscillation cycles \citep[e.g.,][]{2002ApJ...577..475R, 2019FrASS...6...22P}. The driver of decayless kink oscillations which typically have displacement amplitudes about 1~Mm or smaller, is still debated \citep[e.g.,][]{2016A&A...591L...5N, 2024A&A...688A..80K, 2025A&A...696A.125P}. Other coronal plasma structures have also been observed to exhibit kink oscillations, such as {streamers \citep[e.g.,][]{2010ApJ...714..644C,2022MNRAS.515.4055G}}, prominences \citep[e.g.,][]{2011A&A...531A..53H, 2012LRSP....9....2A}, cavities \citep[e.g.,][]{2018ApJ...860..113Z} and jets \citep[e.g.,][]{2014A&A...562A..98C}. 

The oscillation period of kink oscillations is determined by the plasma density and the magnetic field strength \citep[e.g.,][]{1983SoPh...88..179E}, which makes this coronal wave phenomenon a reliable tool for the estimation of the coronal field \citep[e.g.,][]{2024RvMPP...8...19N}. Other important seismological applications of kink oscillations include the diagnostics of fine perpendicular structuring \citep[e.g.,][]{2003ApJ...598.1375A} and the estimation of the density scale height \citep[e.g.,][]{2009SSRv..149....3A}. Furthermore, the decayless regime is expected to shed light on the energy supply to the corona \citep[e.g.,][]{2017A&A...604A.130K, 2019FrASS...6...38K, 2020SSRv..216..140V, 2023NatCo..14.5298Z}.

So far, the detection of kink oscillations has been {predominantly} performed by visually inspecting movies constructed from data cubes, i.e., time sequences of images. The most comprehensive catalogue of decaying kink oscillations includes only 223 oscillating loops in 96 oscillation events, found during the time span from 2014 May 20 to 2018 December 26 \citep{2019ApJS..241...31N}. A number of oscillation events could have been missing. Furthermore, no comprehensive catalogue of decayless kink oscillations has yet been created, except some limited to several tens of events \citep[e.g.,][]{2015A&A...583A.136A, 2024A&A...690L...8L}. The search for events of interest is strongly complicated by the huge amount of data. For example, only in the 171~\AA\ channel, the Atmospheric Imaging Assembly (AIA) instrument on the Solar Dynamics Observatory (SDO) spacecraft delivers about 7200 $4096\times 4096$ images per day, which corresponds to about 230~GB each day \citep{2012SoPh..275...17L}. Thus, there is a clear need for the use of machine learning (ML) techniques for the detection of kink oscillations in observational data sets delivered by AIA. Such a technique would be also applicable to the search for kink oscillations in data obtained with newer imaging instruments, such as the Extreme Ultraviolet Imager (EUI) on Solar Orbiter \citep{2020A&A...642A...8R}, and the Association of Spacecraft for Polarimetric and Imaging Investigation of the Corona of the Sun (ASPIICS) {on} Proba-3 \citep{2022cosp...44.1326G}. 

The use of ML techniques in solar physics is a rapidly developing research avenue \citep[e.g.,][]{2023LRSP...20....4A}. A number of recent studies address ML approaches to the prediction of flares and eruptions \citep[see, e.g.,][for a recent review]{2025RAA....25c5025H}.
Applications of ML also include forecasting of the solar cycle \citep[e.g.,][]{2022MNRAS.515.5062B}, inferring parameters of photospheric flows \citep[e.g.,][]{2020FrASS...7...25T}, identifying and tracking solar magnetic flux elements and features in observed vector magnetograms \citep[e.g.,][]{2020ApJS..250....5J}, creating {farside magnetograms \citep[e.g.,][]{2019NatAs...3..397K, 2021NatAs...5..108L, 2022ApJS..262...50J},} reconstructing the 3D geometry of coronal loops from EUV 2D images \citep[e.g.,][]{2021ApJ...910L..10C}, automated detection and analysis of coronal active-region structures \citep[e.g.,][]{2024MNRAS.532..965G}, spectropolarimetric inversions \citep[e.g.,][]{2020A&A...644A.129M}, image enhancement \citep[e.g.,][]{2018A&A...614A...5D}, and many other tasks \citep[e.g.,][]{2024SDIS....190688C}.
However, the potential of ML has not yet been fully exploited in coronal oscillation studies. As a first step in that direction, \citet{2024ApJS..274...31B} developed a fully convolutional neural network trained to detect rapidly decaying harmonic oscillatory patterns in light curves of solar and stellar flares. Another convolutional neural network has been designed and trained to automatically detect filament oscillations
\citep{2025A&A...694A.237C}. Both studies have demonstrated the applicability of ML to the detection of oscillatory patterns in imaging data sets. 

The aim of this paper is to develop a neural network for the automated detection of kink oscillations of coronal loops in 3D (two spatial coordinates and time) imaging data cubes. The paper is organised as follows. In Section~\ref{sec:data}, we describe the datasets we use to train and validate the neural network developed. The description of the model architecture and the assessment of its performance can be found in Section~\ref{sec:model}. Next, in Section~\ref{sec:install}, a browser application of the developed model is discussed. Finally, in Section~\ref{sec:conc}, we summarise our results and discuss future prospects.  

\section{Dataset} \label{sec:data}
\subsection{Synthetic dataset}\label{sec:synth_data}
Deep learning models typically require large amounts of labeled data for both training and validation. However, in the existing catalogue \citep{2019ApJS..241...31N}, there are less than 200 kink oscillation events which exhibit signals clear enough for direct use in training. To address this shortage, we generated a comprehensive synthetic set of kink oscillation data cubes, i.e., digital movies, to serve as the source of the first training phase of our network.

\begin{figure}[h]
\gridline{\fig{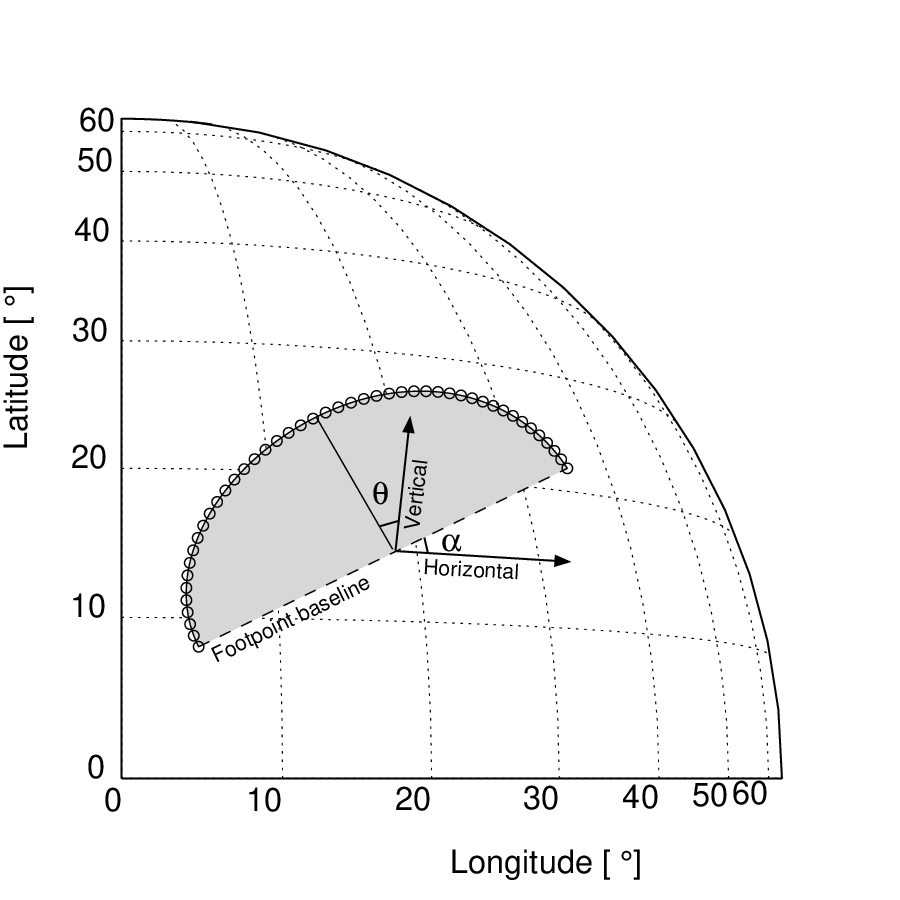}{0.45\textwidth}{(a)}
          \fig{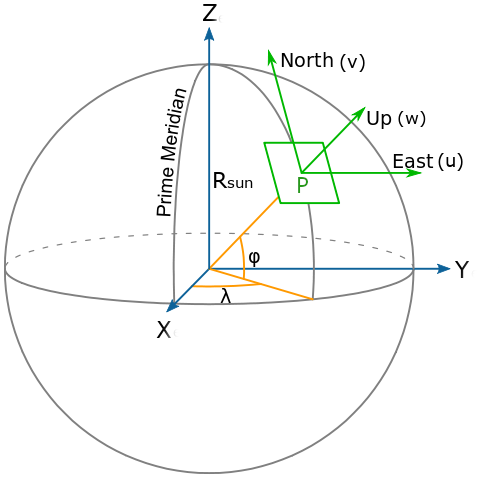}{0.45\textwidth}{(b)}}
  \caption{(a) {Definition of the 3D loop coordinate system in the loop plane (grey) with respect to heliographic coordinates. Here, $\alpha$ denotes the azimuth angle between the footpoint baseline and the heliographic east–west direction, and $\theta$ is the inclination angle between the loop plane and the local vertical to the solar surface. The hollow circles represent the boundary of the local coronal loop cross-section, which together form a semi-torus.} (b) Definition of the spherical coordinate system ($\lambda, \varphi, \mathrm{R_{sun}}$) and the local ENU coordinate system ($u, v, w$) of the loop plane with respect to the Heliocentric Earth Equatorial (HEEQ) coordinates ($x, y, z$).}
  \label{fig:geoloop}
\end{figure}    

To generate the synthetic dataset, we first constructed a three-dimensional geometric model of the solar corona populated with multiple coronal loops. On the visible side of the solar hemisphere, we defined 10,000 regions of interest (RoIs), each measuring 
$600 \times 600$~arcsec, and positioned them randomly within 
$\pm\ 60^\circ$ of solar latitude.
Coronal loops often appear in clusters because they originate from localized low-atmospheric magnetic sources, i.e., typically anchored at footpoints concentrated around regions of high magnetic flux (e.g., sunspots and pores). Hence, in the heliographic coordinates, we randomly selected between 1 and 3 points within each RoI as cluster centers. For each cluster center, 2 to 4 points were randomly distributed in its vicinity to serve as a single footpoint of each synthetic coronal loop. Consequently, each simulated RoI comprises between 2 and 12 loops. The planes of these loops were assigned through a fully random azimuthal orientation $\alpha$ and an inclination angle $\theta$ within $\pm 45^\circ $ (see Figure~\ref{fig:geoloop}a) on the spherical surface.  

Following the geometrical model {used} by \cite{2023NatCo..14.5298Z}, we then modelled coronal loops as semi-tori in the local ENU Cartesian coordinate system (Figure~\ref{fig:geoloop}b) as 
\begin{align}
   \label{eq:psta}
     &u = (R + r \cos{\nu}) \cos{\phi},  \\
     &v = (R + r \cos{\nu}) \sin{\phi}, \nonumber \\
     &w= r \sin{\nu},\nonumber
\end{align}
where $R$ and $r$ are the major and minor radii of the loop, respectively. $\phi \in [0,\pi]$ {measures the toroidal position along the major radius of the torus}, and $\nu \in[0,2\pi]$ {measures the poloidal direction around the cross-sectional circle of the minor radius}. Since the polarization of oscillations may depend on the direction of the external driving force, we considered five distinct types of polarization. One of the following five polarisation types was randomly assigned to each oscillation: (1) horizontal polarization, (2) vertical polarization, (3) oblique polarization, (4) circular polarization, and (5) elliptical polarization. The loop {axis} displacements were implemented as
\begin{align}
  \label{eq:posc}
  \text{Horizontal: }&r = 1 + A_0 \sin{\phi} \sin{(m\nu)} \sin{(\omega t)},  \\
  \text{Vertical:   }&r = 1 + A_0 \sin{\phi} \cos{(m\nu)} \sin{(\omega t)}, \nonumber
\end{align}
where $A_0$ is the perturbation amplitude, $m=1$ indicates the kink symmetry of the perturbation, and $\omega t$ is the oscillation phase in radians. 

In 5,000 simulated RoI, a random number of coronal loops (at least one per region) were initialized with oscillatory motions (Eq.~\ref{eq:posc}) to produce positive samples{, with at least one stationary loop retained in each positive sample}. The remaining 5,000 regions contained only stationary loops (Eq.~\ref{eq:psta}), i.e., negative examples. Subsequently, we adopted an appropriate time resolution, e.g., 12 seconds consistent with the AIA cadence, and transformed the loop coordinates in each frame back into the heliocentric coordinate system ($x$, $y$, $z$) using the following relation, with reference to the coordinates of the point $P=(P_x, P_y, P_z)$:
    \begin{equation}
    \begin{bmatrix}
    x \\ y \\ z
    \end{bmatrix}
    =
    \begin{bmatrix}
    -\sin(\lambda) & -\sin(\varphi)\cos(\lambda) & \cos(\varphi)\cos(\lambda) \\
    \cos(\lambda) & -\sin(\varphi)\sin(\lambda) & \cos(\varphi)\sin(\lambda) \\
    0 & \cos(\varphi) & \sin(\varphi)
    \end{bmatrix}
    \begin{bmatrix}
   u \\ v \\ w
   \end{bmatrix} 
   +
   \begin{bmatrix}
   P_x \\ P_y \\ P_z
   \end{bmatrix}.
   \label{eq:trans}
   \end{equation}

The next step is to generate synthetic intensity maps. We projected the 3D coronal loops ($x, y, z$) in each frame onto the $y$-$z$ plane (Figure~\ref{fig:geoloop}b) and interpolated the synthetic emission onto a field-of-view grid with $0.6$~arcsec resolution, imitating AIA's resolution, assigning integer intensity values. This spatial resolution could be moderately reduced to shorten the data generation time, and increased to mimic a data set obtained with a higher resolution coronal imager, e.g., EUI. Thereafter, a low background intensity variation was added, with both red noise and Poisson noise being superimposed across the entire RoI.

\begin{figure}
    \centering
    \includegraphics[width=0.75\linewidth, trim={0 0 0 430}, clip]{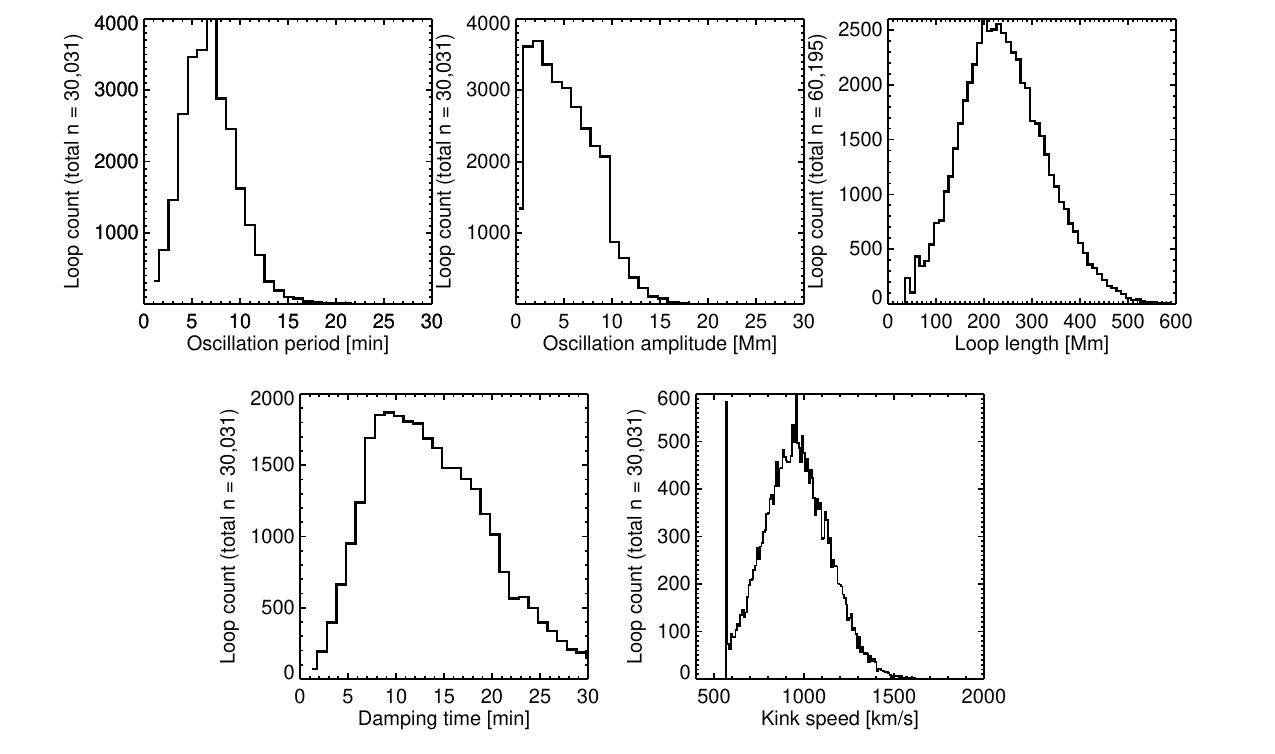}
    \caption{Parameter distributions of the synthetic coronal loops. A total of 60,195 coronal loops are embedded across the 10,000 synthetic events. The distributions are consistent with observed distributions \citep{2019ApJS..241...31N}.}
    \label{fig:sypara}
\end{figure}

The key parameters of the kink oscillations, used in the simulations, are taken from the catalogues of \citet{2016A&A...585A.137G, 2019ApJS..241...31N}. We first assign the loop length $L$, and the kink speed $C_\mathrm{k}$ according to Gaussian distributions centered at $285\ \mathrm{Mm}$ and $1000\ \mathrm{km/s}$, respectively. The oscillation period can then be calculated using $P = 2L / C_\mathrm{k}$. The oscillation amplitude $A_0$ is sampled from a Gaussian distribution within the range of $0.5 \% $ to $5 \%$ of the loop length. Considering projection effects, the actual amplitude may be larger. The minor radii $r$ of the loops are randomly drawn from a uniform distribution between 1~Mm and 3~Mm. Although resonant absorption predicts a damping time $\tau_d$ proportional to the period, observations also suggest an amplitude-dependence, indicating possible nonlinear effects \citep{2016A&A...590L...5G, 2019ApJS..241...31N}. Hence, an empirical relation $\log (\tau_d / P) < -0.7\ \log(A_0) +2.8 $ is adopted as the upper limit of the damping time. The distributions of the parameters of synthetic coronal loops, generated in our study, are summarized in Figure~\ref{fig:sypara}. Finally, to ensure that the network can fully capture each oscillatory motion, we set the movie duration to exceed three times the oscillation period. For negative samples, the durations are randomly assigned to be tens of minutes. The dataset has been uploaded to Harvard Dataverse \citep{Dataset1}.


\subsection{Real dataset}\label{sec:real_data}
We employed SDO/AIA 171~\AA\ channel image sequences as real training samples, exploiting their long-term, continuous, high-resolution full-disk coverage and the extensive SDO archival database. {As the synthetic data set, the real data set used in this study comprises two subsets too. Some loops exhibit kink oscillations coexisting with non-oscillating loops, while others contain no oscillating loops at all. These subsets are labeled as positive (“P”) and negative (“N”), respectively.} Each sample was defined by a region of interest of $260 \times 260$ pixels over a 30-minute duration, allowing a focus on the target structures while limiting data storage requirements. We preserved the original $0.6$ arcsec resolution and produced high-quality animations from the image sequences at 15 frames per second as the sample format. The entire real database comprises 50 positive samples and 128 negative samples, which occupies approximately 600~MB of storage.

\begin{figure}
    \centering
    \includegraphics[width=0.65\linewidth, trim={0 0 300 480}, clip]{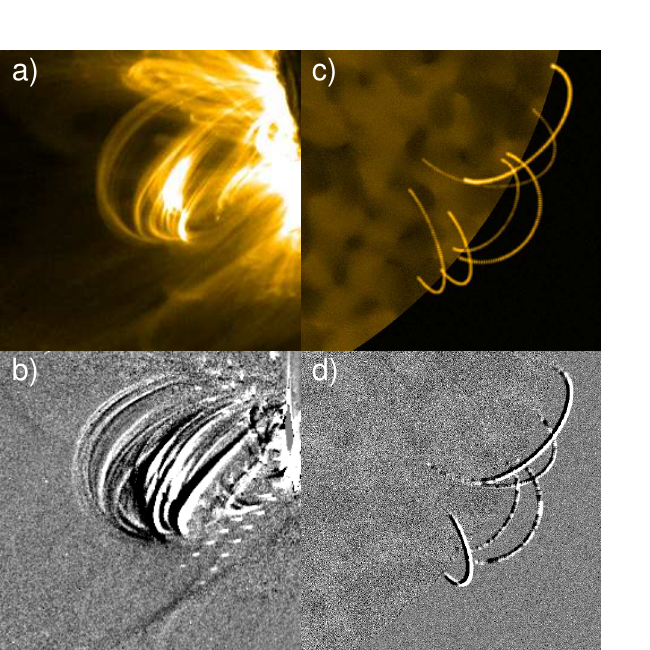}
    \caption{Frames from real (a) and synthetic (c) data samples experiencing kink oscillations. Panels (b) and (d) show the running differences of the samples presented in panels (a) and (c), respectively. In the running difference images, the kink oscillations are clearly evident as black-and-white stripes highlighting the oscillating loops.}
    \label{fig:datasetcomp}
\end{figure}

Most of the positive examples, i.e., those with oscillations, were selected from \cite{2019ApJS..241...31N}, with some supplementary events from an extended catalogue (Zhong et. al, in prep.) occurring between 2022 and 2024. Three main criteria were applied in selecting events: 1) {large-amplitude, clearly damped oscillations excited by a disturbance from a nearby eruption}, to exclude decayless cases; 2) clearly identifiable loops or loop bundles with {well-defined boundaries and/or loop axes}; and 3) at least three complete oscillation cycles observed within the observational time interval. In contrast, clearly identifiable stationary loops are far more common than kink-oscillating loops. At both the solar limb and on the disk, we collected a substantial number of samples, covering coronal loops in both active regions and the quiet Sun. This dataset is exposed in Harvard Dataverse \citep{Dataset2}. Positive examples from both datasets are shown in Figure \ref{fig:datasetcomp}

\section{The model and its performance} \label{sec:model}
\subsection{The model architecture}
\label{sec:architecture}

\begin{figure}[h]
\includegraphics[width=\textwidth]{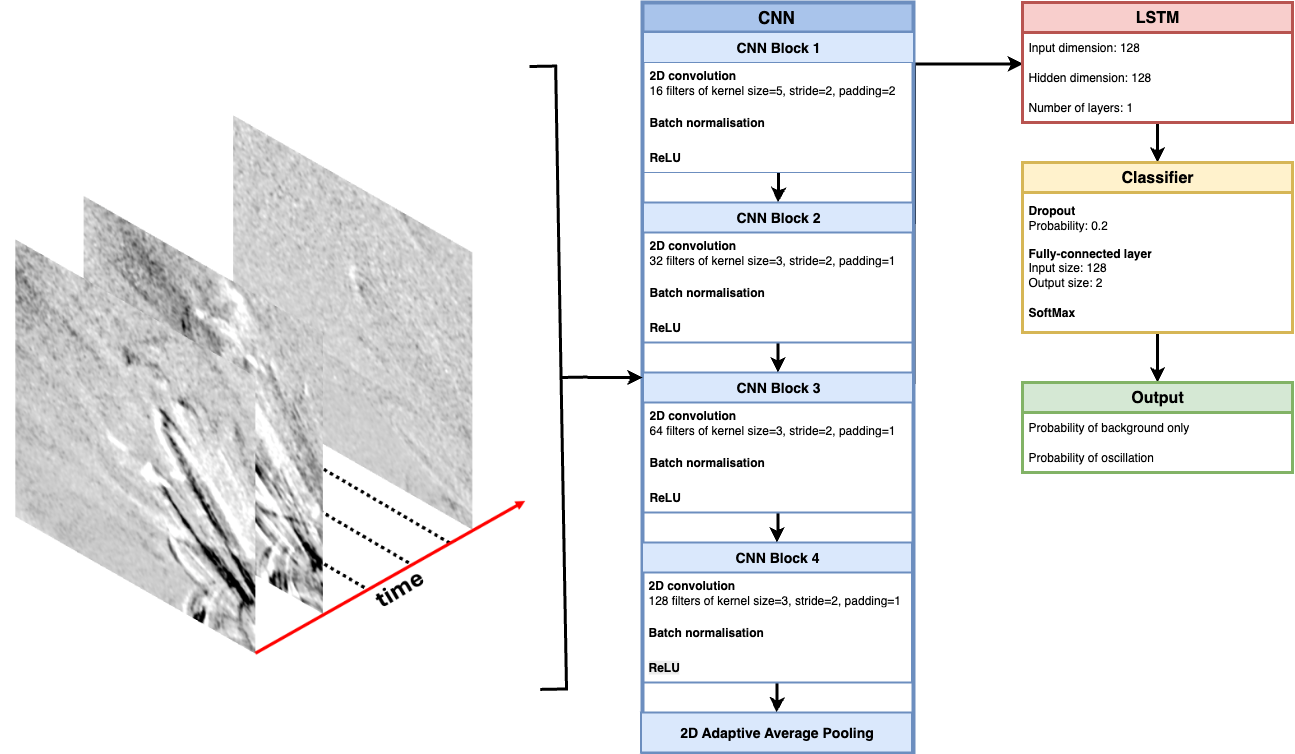}
\caption{Architecture of the CNN-LSTM network for a kink oscillation detection task.}
\label{fig:architecture}
\end{figure}

Detection of kink oscillations in the corona can be classified as an action recognition task. For this type of task, both spatial and temporal information are crucial. {A possible way} to solve this problem is to add an additional temporal dimension to usual Convolutional Neural Networks (CNN) to process a full data cube (two spatial dimensions plus time). This type of model is called 3D CNN and was proposed for the first time in \citet{Ji2013}. Despite their {effectiveness}, 3D CNNs are computationally heavy, and other hardware-friendly architectures can be used. As an alternative, a CNN can be stacked with a network processing sequences, for example, with a Recurrent Neural Network (RNN). In this case, a CNN processes an original video frame-by-frame and creates a sequence of lower-dimensional feature vectors instead of a frame sequence. Next, these feature vectors are fed into an RNN which processes the data along the temporal dimension. The example of this architecture is CNN-LSTM proposed by \citet{Donahue2015}.

In this study, we use a MacBook Pro M3 with 8GB of shared memory. Due to memory constraints, we opted for the CNN-LSTM architecture over 3D CNNs, as it is less demanding in terms of GPU and RAM usage. Our model architecture is shown in Figure~\ref{fig:architecture}. Rather than using a state-of-the-art (SoTA) pre-trained model such as ResNet \citep{He2016} to extract feature vectors from individual video frames, we created a shallow custom CNN backbone consisting of four convolutional layers followed by a 2D adaptive average pooling layer. We prefer a shallow custom backbone over a SoTA model for several reasons. First of all, ResNet and other SoTA architectures are typically trained on the ImageNet dataset, which contains natural images (e.g. animals, objects) that differ substantially from solar coronal images. Next, even if we assume that the SoTA models can extract relevant features from solar images, a shallow custom backbone allows for much faster inference and efficient training from scratch on our relatively small, task-specific dataset. Thus, a shallow CNN backbone is potentially capable of learning a small set of domain-related features.

The sequence of frame-level features produced by the CNN is passed to an LSTM module to analyse the temporal dynamics. The LSTM updates its internal state step by step, and the final hidden state is passed to a fully connected layer that outputs a two-dimensional probability vector. The first value represents the probability that the video does not contain kink oscillations; the second value indicates the probability of their presence.

Before being fed into the network, each video undergoes the following preprocessing steps:
\begin{enumerate}
    \item Image-differencing, $\mathrm{I}_{t,m,n}^d=\mathrm{I}_{t,m,n}-\mathrm{I}_{t-1,m,n}$, where $\mathrm{I}_{t,m,n}$ is the value of intensity in a pixel with coordinates $m$ and $n$ at time $t$. This operation helps to highlight motions on a video sample.
    \item {Clipping values below the 0.01 percentile and above the 99.99 percentile, computed over the entire video, to mitigate flickering between frames.}
    \item Min-max scaling to have the video data in range (0, 1), $\mathrm{I}_{t,m,n}^{dm}=\left(\mathrm{I}_{t,m,n}^d-\min\limits_{t, m, n}\mathrm{I}_{t,m,n}^{d}\right)/\left(\max\limits_{t, m, n}\mathrm{I}_{t,m,n}^{d}-\min\limits_{t, m, n}\mathrm{I}_{t,m,n}^{d}\right)$.
    \item Standardising with ImageNet statistics to stabilise training, $\mathrm{I}_{t,m,n}^{dms}=\left(\mathrm{I}_{t,m,n}^{dm}-0.485\right)/0.229$. This statistics showed better training stability compared to the statistics calculated for the training synthetic dataset.
    \item Rescaling to the $250\times250$ frame size.
\end{enumerate}

\subsection{Performance on synthetic data}
\label{sec:synth_performance}

\begin{figure}[h]
\includegraphics[width=\textwidth]{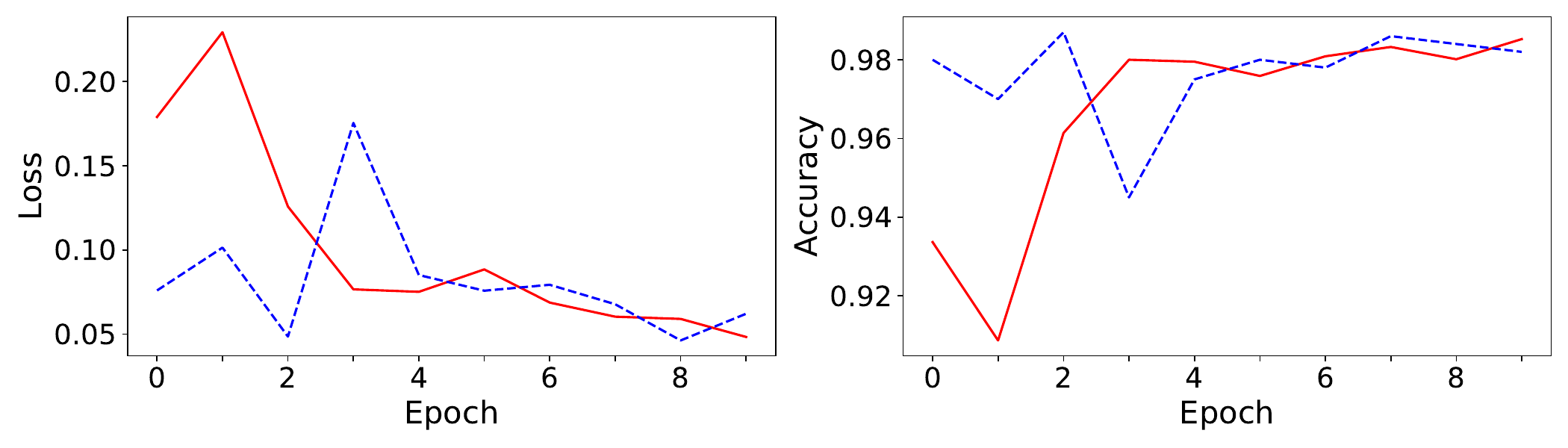}
\caption{Left panel: the loss function during the training process for the {training} (red curve) and validation (blue dashed curve) datasets. Right panel: the accuracy score on the {training} (red curve) and validation (blue dashed curve) datasets during the training process.}
\label{fig:train_curves}
\end{figure}

\begin{figure}[h]
\includegraphics[width=\textwidth]{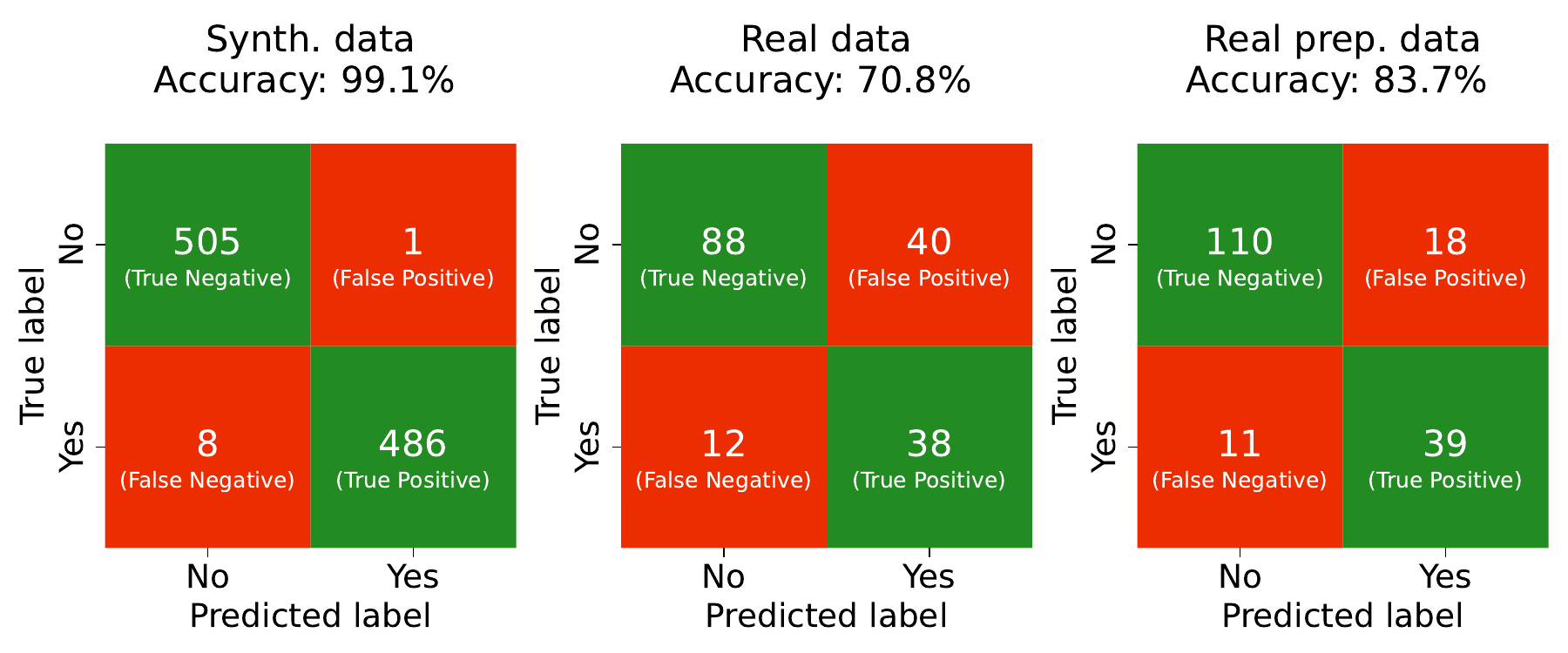}
\caption{Confusion matrices for the synthetic test dataset (left panel), the real dataset (middle panel), and the preprocessed real dataset (right panel).}
\label{fig:cm}
\end{figure}

Due to hardware limitations, the network was trained using small batches of two movies, with gradient accumulation over eight steps, resulting in an effective batch size of 16. The learning rate was set to $10^{-3}$. Although LSTMs are agnostic to the input sequence length, batch training requires that all sequences within a batch have the same shape. To address this, we standardise the movie length to 100 frames after preprocessing: movies longer than 100 frames are truncated at the end, while shorter ones are padded by repeating their first frames until they reach the target length.

Figure~\ref{fig:train_curves} shows the evolution of the loss and accuracy for the training and validation datasets during the training process. It can be seen from the right panel of Figure~\ref{fig:train_curves} that the accuracy reaches 98\% for both the training and validation datasets. Based on the validation loss curve, we chose the model at epoch 8 as the final model to test.

The left panel of Figure~\ref{fig:cm} shows the confusion matrix calculated for the synthetic test dataset, and the accuracy  reaches on this dataset 99.1\% with a good balance between false positive and false negative detections.

\subsection{Performance on real data}
\label{sec:real_performance}
To assess the model performance on the real data, we use the data described in Section~\ref{sec:real_data}. The confusion matrix calculated for this data (the middle panel of Figure~\ref{fig:cm}) shows that the model performs worse on this real-world data, showing the accuracy of about 70.8\%. This degradation in the model performance is expected because real and synthetic datasets may have different dynamical features and statistical properties. For example, real samples can have a lower contrast due to faint loops, more complicated background dynamics, instrumental noise, other types of wave motions, etc. In general, it is currently impossible to reproduce the actual morphology and dynamics of the solar corona because of the lack of understanding of the processes which control it. This problem is known as a domain gap, and there are many techniques to mitigate it, including a more careful dataset engineering, prepocessing techniques to make the domains closer, and special model architectures. This issue should be addressed in a follow up study.

\begin{figure}[h]
\includegraphics[width=\textwidth]{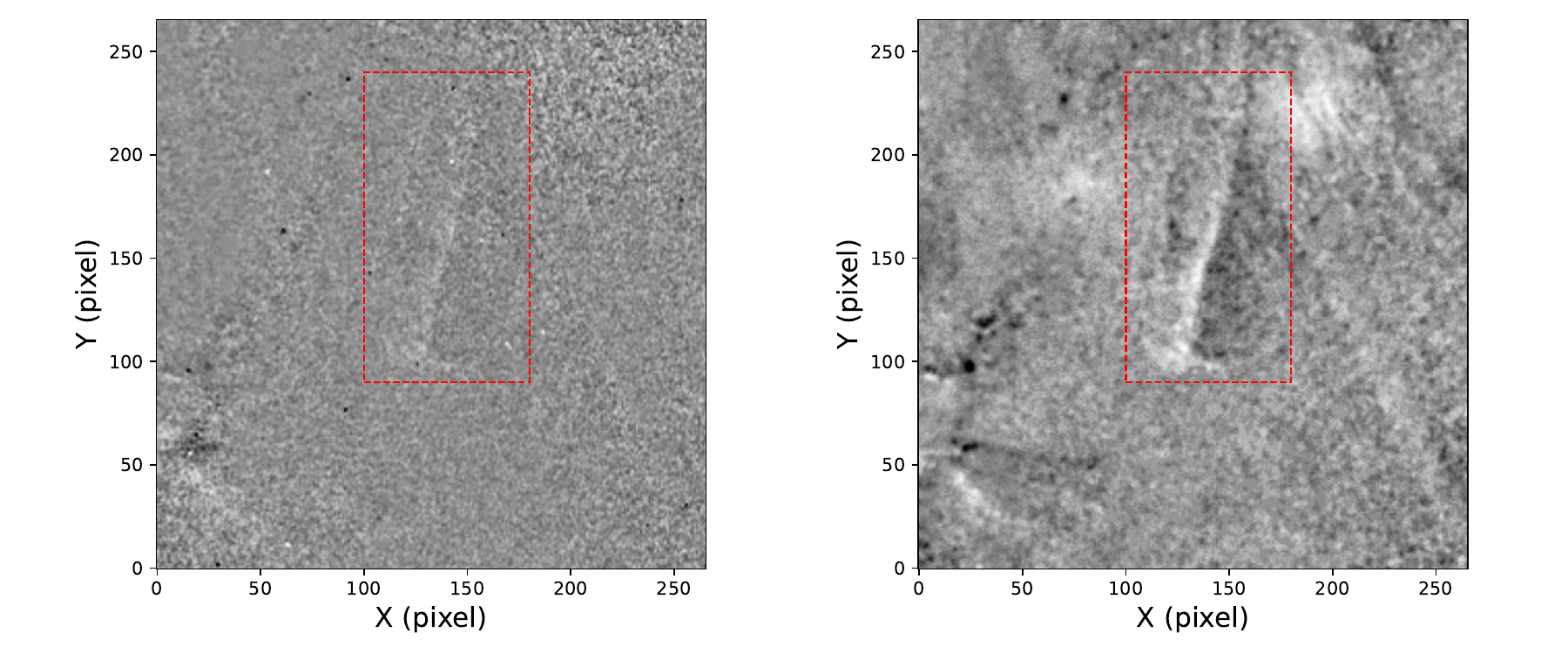}
\caption{Left panel: a frame from a running image difference movie of a kink oscillation. Right panel: the same frame but after preprocessing. In both the panels, the {dashed red rectangle shows the location of the oscillating loop}.}
\label{fig:prep}
\end{figure}

To reduce the domain gap, in this study, we employ a simple preprocessing routine for the real dataset before {applying} the processing routine described in Section~\ref{sec:architecture}. This preprocessing routine includes three steps:
\begin{enumerate}
    \item Median filtering the pixel values along the temporal axis with a 3-point filter to reduce photon noise.
    \item Gaussian smoothing with $\sigma=1$ applied to each frame to smooth {short-scale} spatial noise.
    \item  The Contrast Limited Adaptive Histogram Equalization (CLAHE) procedure was applied to each frame to enhance its contrast.
\end{enumerate}
With this preprocessing, the accuracy on the real test data increased to 83.7\%. At the same time, a false positive number dropped from 40 to 18 (see the right panel of Figure~\ref{fig:cm}). Thus, this performance increase suggests that this simple preprocessing routine allows us to bring the data domains closer. Figure~\ref{fig:prep} compares {the cases with and without preprocessing}. It can be seen that preprocessing enhances the oscillation (black and white stripes highlighting the oscillating loop) and makes it easier to detect.

It can be instructive to examine the embedding space where the network maps its input. Since we use a CNN-LSTM architecture to process data cubes, the CNN and LSTM blocks sequentially process the model input. The CNN block transforms each $250\times250$ frame into a set of 128 elements, i.e, a 128-dimensional feature vector. Frames that are \lq\lq semantically\rq\rq\ similar (e.g., showing {similar spatial structures}) are expected to be mapped to nearby points in this space, while frames with distinct characteristics should be mapped further apart \footnote{For example, in natural language processing, the distance in a multidimensional feature space between the words "candy\rq\rq\ and "cake\rq\rq\ should be smaller than the distance between either of them and the word "tree".}. The series of vectors produced by the CNN block is passed to the LSTM block which captures the temporal dynamics.{ The LSTM output is also a 128-dimensional vector. Unlike the CNN features, here a single vector represents an entire video sample.} Thus, for an input movie of size $100\times250\times250$, the CNN outputs a series of vectors with a total size of $100\times128$, which are then compressed by the LSTM block into {a single 128-dimensional feature vector}. This final vector reflects how the model represents the semantics of its input. Consequently, frames containing kink oscillations should form a distinguishable group from the frames without oscillations.

\begin{figure}[h]
\includegraphics[width=\textwidth]{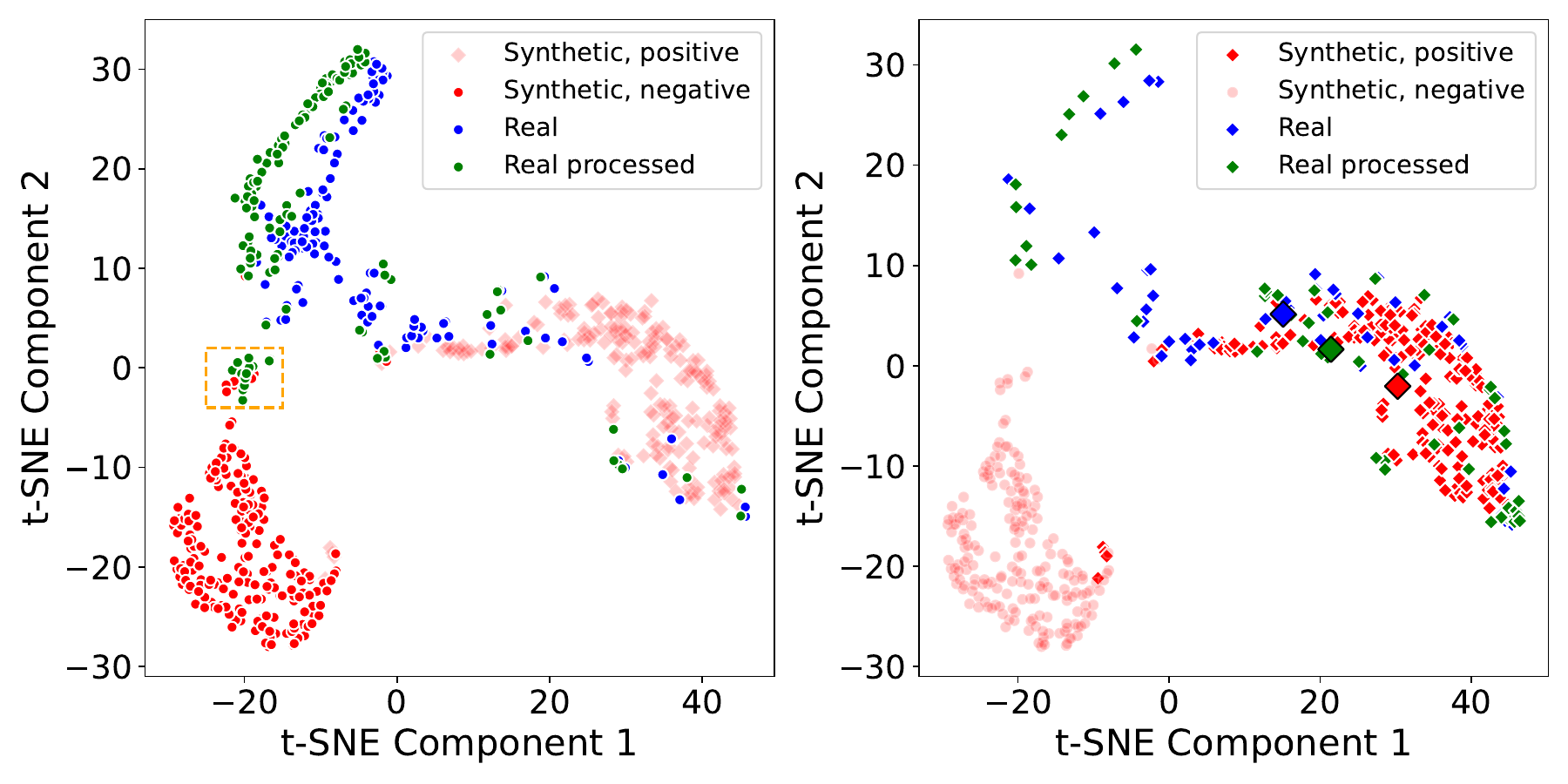}
\caption{The t-SNE visualization of LSTM feature embeddings for 200 randomly selected positive and 200 negative synthetic samples combined with all real samples. Diamonds and circles denote samples with and without {oscillations}, respectively. Left panel: relationship between  real data samples without oscillations and the clustered synthetic data. Right panel: relationship between real data samples with oscillations and the clustered synthetic data. Large markers indicate the centres of mass of the data clouds. {Colour coding for synthetic samples was adjusted to emphasise negatives in the left panel and positives in the right.}}
\label{fig:emb_lstm}
\end{figure}

To visualise this latent space, we randomly selected 200 positive and 200 negative synthetic samples and included all available real data. For each movie, we extracted the LSTM output vector, and then applied t-distributed Stochastic Neighbor Embedding (t-SNE)  \citep{maaten2008visualizing} to reduce the 128-dimensional vectors to two components, allowing 2D visualisation. The result is shown in Figure~\ref{fig:emb_lstm}. In both panels, diamonds and circles represent samples with and without oscillations, respectively. 
It is visible that the synthetic data form two distinct clusters: the left cluster consisting mainly of negative samples and the right cluster dominated by positive samples, i.e., with kink oscillations. The left panel of Figure~\ref{fig:emb_lstm} shows how real samples without oscillations are distributed before and after preprocessing, compared with synthetic samples. The real samples form a separate cluster and have a small overlap with the positive synthetic cluster. This indicates that the negative synthetic data fail to fully represent the {real coronal variability}, which often includes complex dynamics such {as other wave and irregular motions, sporadic brightenings,} and other phenomena. However, after preprocessing, {a number of real samples overlap} with the negative cluster (see the dashed orange rectangle) and move away from the positive synthetic cluster. 

The right panel of Figure~\ref{fig:emb_lstm} shows similar data distributions but for real samples containing oscillations. Notably, there is significant overlap between the real samples and the positive synthetic cluster, but no overlap with the negative cluster. It suggests that the synthetic dataset effectively captures the oscillatory patterns present in the real-world data. Although preprocessing does not show an obvious effect on individual points, it becomes apparent when considering the data cloud centers: preprocessing shifts the positive real samples closer to the positive synthetic cluster. Thus, for both negative and positive real samples, preprocessing  moderately helps bridge the domain gap between real and synthetic data, resulting in a 13\% increase in the classification accuracy.

\section{The model distributive} \label{sec:install}
The source code of the model is openly available via a GitHub repository\footnote{\url{https://github.com/Warwick-Solar/Kink-Oscillation-Detector}} {(the frozen stable version of the code is also available via Zenodo \citep{belov_zenodo})}. Within the \texttt{Notebooks} folder of the GitHub repository, a Jupyter notebook is provided to generate a train/test/validation split of the dataset and to train or fine-tune the network. The main component of the repository is a Streamlit-based web application, developed by us to run the trained model and visualise the input and output data.

To install and run the application, the following steps should be pursued (written in the Unix shell syntax and assuming that Anaconda or Miniconda version 24.1.2 or later is pre-installed):

\begin{enumerate}
\item Clone or download the repository from GitHub.
\item In the terminal, navigate to the root directory of the project:\
\verb|cd Kink-Oscillation-Detector| (or \verb|cd Kink-Oscillation-Detector-main|).
\item Create a new Conda environment using the provided configuration file:\
\verb|conda env create -f ./Environment/env.yaml|
\item Activate the newly created environment:\
\verb|conda activate kink_osc_detector|
\item Launch the application\footnote{If any issues occur, consider updating \texttt{streamlit} or \texttt{pytorch-lightning} using:
\texttt{conda update <library\_name>}}:
\verb|streamlit run ./Application/app.py|
\item If the application does not open automatically, navigate to the following URL in your web browser:\
\url{http://localhost:8501} (or another address provided in the terminal output).
\item Use the graphical interface to analyse movies. The input format of the movie should be \texttt{.mp4}. Example movies are available in the \lq\lq Test data\rq\rq\ folder within the project directory.
\end{enumerate}

Steps 1--3 are required only once for the installation. To run the application after the installation, steps 1 (if not already in the project directory) and 4--7 must be repeated. On Windows systems, all commands must be executed via the Anaconda Prompt, not the standard Windows shell.

\begin{figure}[h]
\centering
\includegraphics[width=15cm]{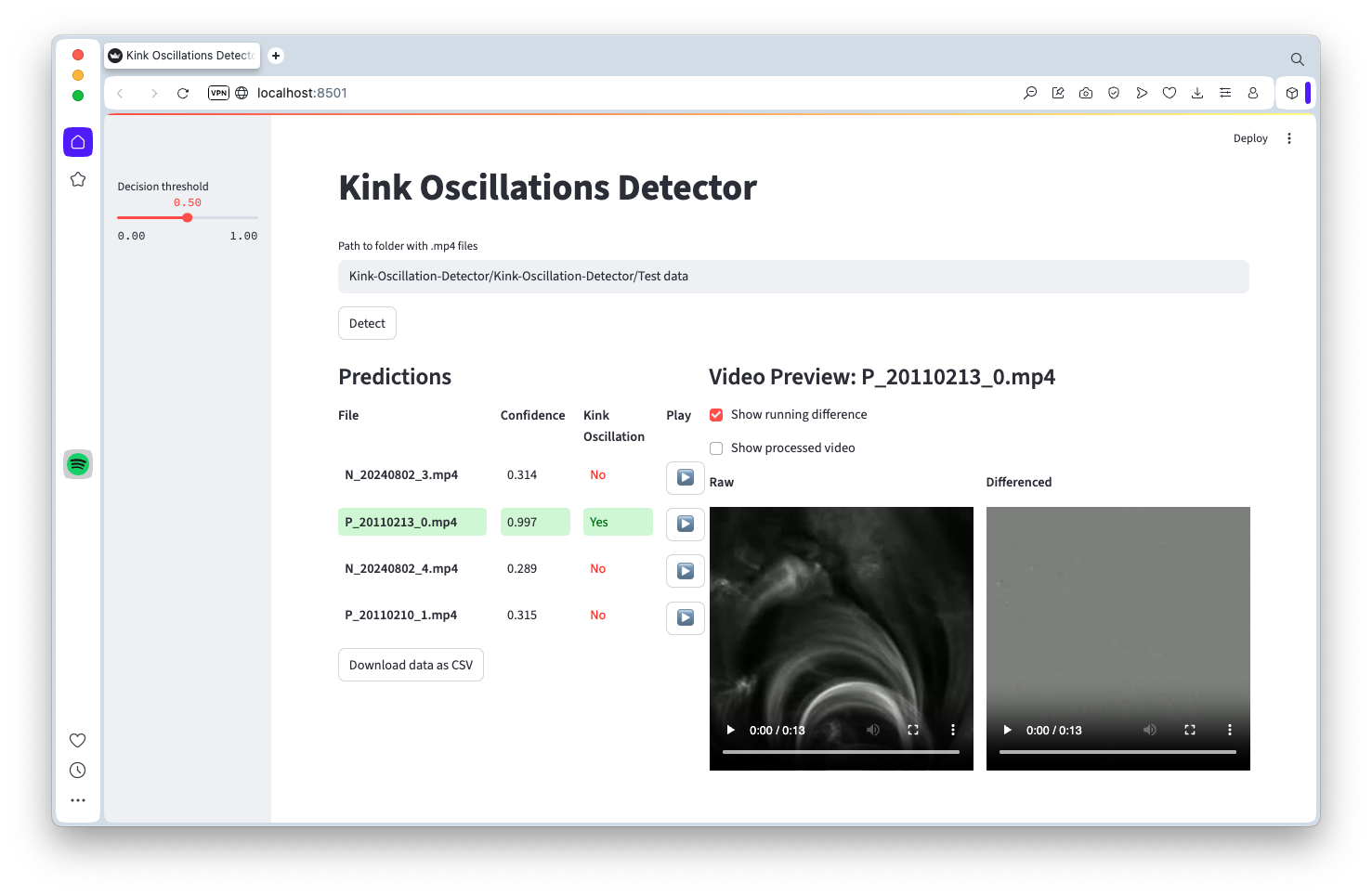}
\caption{The user interface of the browser application for the CNN-LSTM based kink-oscillation detection in solar coronal imaging data.
\label{fig:window}}
\end{figure}

Upon completing steps 4--6, the user should see the application interface in a browser window (see Figure~\ref{fig:window}). Initially, the interface includes a search bar to specify the folder containing the imaging data, the \texttt{Detect} button, and a slider to adjust the detection threshold. After selecting the data folder and clicking \texttt{Detect}, the model processes the movies and generates a table displaying the predicted probabilities of kink oscillations in each movie. Each row of the table includes a play button that allows users to view the corresponding movie. Options are also provided to inspect both the differenced and preprocessed versions of the input data. The results can be downloaded as a CSV file by clicking the \lq\lq Download data as CSV\rq\rq\ button, enabling further analysis.

\section{Conclusions and prospects} \label{sec:conc}
In this work, we designed an ML model based on a CNN–LSTM architecture to detect kink oscillations of coronal loops in solar imaging data cubes (digital movies). The model processes spatio-temporal data in two stages: first, a custom CNN backbone extracts spatial features from individual frames; then, an LSTM block analyzes the resulting feature sequences to capture their temporal evolution. To keep the model computationally efficient, we implemented a relatively shallow network with a simple, custom CNN backbone, enabling training on high-spec laptops.

To train the network, we generated 10,000 synthetic data cubes, half containing kink oscillations and half containing non-oscillating loops. This dataset is publicly available via Harvard Dataverse \citep{Dataset1}. We split it into training, validation, and test subsets in an 80\%/10\%/10\% ratio. After training, the network achieved an accuracy greater than 98\% in all synthetic subsets.
When applied to a real dataset \citep{Dataset2}, the network’s performance dropped to 70.8\% due to a domain gap arising from the greater complexity of real-world data. To mitigate this gap and make the real data more similar to the synthetic training set, we implemented a preprocessing pipeline comprising Gaussian smoothing of individual frames, median filtering of pixel values between frames, and CLAHE-based contrast enhancement. This preprocessing improved the accuracy on real data to 83.7\%.
To assess the impact of this preprocessing and examine how the model represents the data, we visualized the latent feature space of the network using the t-SNE technique (Figure~\ref{fig:emb_lstm}). The resulting plot revealed a clear domain gap between real and synthetic data and confirmed that preprocessing moderately reduced this gap.

To make the model available to the community, we developed a browser application that can be found on our GitHub. This application allows users to search for kink oscillations in individual data cubes provided as MP4 movies and to visualise the results.

Despite its limitations and the established domain gap, the proposed model is a first step towards utilising deep learning techniques to identify coronal loop oscillations. This model can be used as a starting point for developing more robust tools for detecting transverse oscillations in solar coronal imaging data. This can be achieved by applying the network to large volumes of solar data and collecting newly detected events, which should be verified under human supervision to ensure that they are genuine. Such a collection would help expand our existing catalogues of kink oscillations in the solar corona, which is highly demanded for MHD seismology and coronal heating efforts. 

Future progress could be achieved by adding new synthetic features that better reflect the dynamics of the corona or by combining synthetic and real features within a single sample. An alternative approach is to build a composite dataset containing both real and synthetic samples.

In addition to improving data quality, future work could explore more complex models, such as SoTA CNN backbones or more advanced architectures like transformers for time-sequence analysis. Furthermore, Domain-Adversarial Neural Networks \citep{ganin2016domain} could be employed to bridge the domain gap between real and synthetic data.

In conclusion, we plan to use the developed network as a tool for coronal data mining {-- effectively extracting subsets of specific events from large-scale observational data volumes.}

\begin{acknowledgments}
The work is supported by the STFC Grant ST/X000915/1 and the Latvian Science Council Grant lzp-2024/1-0023. DYK also thanks the UKRI Stephen
Hawking Fellowship EP/Z535473/1. 
\end{acknowledgments}

\software{NumPy, a Python package for fundamental scientific computing \citep{harris2020array};
SciPy, a Python package for fundamental algorithms for scientific computing \citep{2020SciPy-NMeth}; opencv-python, a computer vision library pre-built for Python;
PyTorch Lightning, the deep learning framework;
Streamlit, a Python library for deploying ML projects.}

\bibliography{refs}{}
\bibliographystyle{aasjournal}



\end{document}